\documentclass[aps,pre,reprint,showpacs,notitlepage,onecolumn]{revtex4-1}
\usepackage{epsfig,graphics,amssymb,amsmath,subeqnarray,color,bm}
\usepackage[toc,page]{appendix}
\usepackage{mathtools}
\usepackage{xcolor}
\usepackage{float}

\numberwithin{equation}{section}
\usepackage{graphicx}
\usepackage[colorlinks=true,linkcolor=blue]{hyperref}

\begin{document}
\title{An elastic two-sphere swimmer in Stokes flow}
\author{Babak Nasouri}
\affiliation{
Department of Mechanical Engineering,  
University of British Columbia,
Vancouver, B.C., V6T 1Z4, Canada}
\author{Aditi Khot}
\affiliation{
Department of Mechanical Engineering,  
University of British Columbia,
Vancouver, B.C., V6T 1Z4, Canada}

\author{Gwynn J. Elfring}\email{Electronic mail: gelfring@mech.ubc.ca}
\affiliation{
Department of Mechanical Engineering,  
University of British Columbia,
Vancouver, B.C., V6T 1Z4, Canada}

\begin{abstract} 
Swimming at low Reynolds number in Newtonian fluids is only possible through non-reciprocal body deformations due to the kinematic reversibility of the Stokes equations. We consider here a model swimmer consisting of two linked spheres, wherein one sphere is rigid and the other an incompressible neo-Hookean solid. The two spheres are connected by a rod which changes its length periodically. We show that the deformations of the body are non-reciprocal despite the reversible actuation and hence, the elastic two-sphere swimmer propels forward. Our results indicate that even weak elastic deformations of a body can affect locomotion and may be exploited in designing artificial microswimmers.
\end{abstract}
\maketitle
\section{Introduction}
\label{intro}
In the microscale realm of motile cells, inertia is unimportant and the effect of viscous dissipation dominates the fluid forces on swimming bodies \cite{happel1981,kim1991}. To propel forward in this regime, many microorganisms deform their bodies periodically by converting cells' chemical energy into mechanical work \cite{roberts2013}. As a direct consequence of this inertialess environment, to achieve nonzero net locomotion, such body deformations cannot be invariant under time reversal \cite{purcell1977}. This constraint, colloquially referred to as the scallop theorem, indicates that due to the kinematic reversibility of the field equations in the low Reynolds number regime, reciprocal body distortions have no net effect.

Theoretically, the scallop theorem can be eluded under two circumstances: non-reciprocal kinematics or a violation of the theorem's assumptions (see \cite{lauga2011} and the references therein). The latter exploits the fact that the scallop theorem is solely valid for inertialess single swimmers in quiescent viscous fluid. Therefore, hydrodynamic interactions \cite{trouilloud2008}, a non-Newtonian medium \cite{lauga2009b}, or inertia \cite{gonzalez-rodriguez2009} can all lead to propulsion. Non-reciprocal kinematics are employed by many motile cells in nature to facilitate motion \cite{lighthill1975,lauga2009a}, and also become a key design principle for model swimmers at small scales. In 1977, Purcell introduced a simple three-link swimmer with two rotational hinges that can change its shape in a non-reciprocal fashion, leading to a locomotion \cite{purcell1977}. Subsequently, several analytical model swimmers have been devised wherein non-reciprocal shape change provides the propulsive thrust \cite{dreyfus2005,iima2009,golestanian2010,najafi2010}. Notably, \citet{najafi2004} proposed a simple three-sphere swimmer, in which spheres are identical and connected by two slender rods. The connecting rods change their length in a four-stage cycle that is not invariant under time reversal. After completion of one cycle, the swimmer recovers its original shape but has been translated forward (see also \cite{golestanian2008} and \cite{leoni2009}). \citet{avron2005} suggested a more efficient, yet as simple, swimmer that consists of two linked spherical bladders of different radii. To compensate for the third sphere, they relaxed the rigidity constraint by allowing instantaneous volume exchange between spherical bladders in each stroke. The shape change of the bladders along with the periodic change in their distance, leads to a net displacement of the swimmer. Inspired by these two models, in this paper we investigate a simple, but less intuitive, two-sphere swimmer where one of the spheres is elastic. We propose that the elastic deformation of the swimmer can be sufficient to escape the scallop theorem, alter hydrodynamic interactions and eventually lead to propulsion.

Elasticity, as an inevitable characteristic of motile cells, can significantly affect the hydrodynamics of a motion. The propulsion of flexible bodies \cite{purcell1977,wiggins1998,lagomarsino2003b}, synchronization of flagella \cite{elfring2011b,goldstein2016} and cilia \cite{niedermayer2008,brumley2012,nasouri2016} through elastohydrodynamic interactions, and reorientation of uni-flagellated bacteria due to buckling of the flagellum \cite{son2013,jawed2015} are well-studied examples of such behaviors. For an elastic body in a flow, the balance of viscous forces, external forces and internal elastic forces causes the body to deform and to alter the surrounding flow field, often in a complex fashion \cite{li2013,gao2011,galstyan2015}. \citet{li2013} reported that for an isolated sedimenting filament, elasticity can destabilize the motion and lead to a substantial buckling. Furthermore, \citet{gao2011} showed that elastic spheres in a shear flow exhibit a `tank-treading' motion wherein the particle shape is at steady state while the material points on the boundary are undergoing a periodic motion. However, though seemingly simple, sedimentation of spherical elastic particles in a viscous fluid is largely unexplored. The most recent, and to the best of our knowledge the only, analysis on sedimentation of elastic spheres dates back to more than three decades ago, when \citet{murata1980} investigated the steady state shape deformation of a compressible, Hookean sphere. Using an asymptotic analysis, it was shown that the elastic sphere settles faster and deforms to a prolate spheroid of a smaller volume. In this work, to further investigate the deformation of elastic spheres, we revisit this sedimentation problem but this time for an incompressible neo-Hookean sphere under a prescribed body force. We asymptotically describe the steady state effects of non-linear elastic deformations on the swimming behaviors of an isolated elastic sphere.

The paper is organized as follows. In section~\ref{single-sphere}, we investigate the translation of a single neo-Hookean sphere in Stokes flow. Using an asymptotic approach, we show that for a given body force, due to deformation, the translational velocity of the elastic sphere is smaller compared to a rigid sphere of the same size. Furthermore, we find that the shape deformation is not front-back symmetric and so neither is the flow field generated in the surrounding fluid. In section~\ref{two-swimmer}, we show that by exploiting this asymmetry, the proposed two-sphere model can indeed swim in a low Reynolds number regime. Finally, in the case where the distance between the spheres is relatively large, we determine the propulsion velocity.

\section{Translation of an elastic sphere}\label{single-sphere}
We begin our analysis with considering the translation of an incompressible isotropic neo-Hookean sphere in an otherwise quiescent viscous fluid. The sphere has radius ${R}_0$ and is driven by body force $\mathbf{f}(t)$. In the fluid domain ($\Omega_{\text{f}}$), the flow field around the sphere is governed by the Stokes equations
\begin{align}
\label{stokes1}
\text{div}~{\boldsymbol{\sigma}}_\text{f}=\boldsymbol{0},\\
\label{stokes2}
\text{div}~{\mathbf{v}}=0,
\end{align}
where ${\mathbf{v}}$ is the fluid velocity and ${\boldsymbol{\sigma}}_{\text{f}}$ is the dynamical stress tensor in the fluid domain defined by the constitutive relation
\begin{align}
{\boldsymbol{\sigma}}_\text{f} = -{p}_\text{f}\mathbf{I} + {\eta}_\text{f}\left[\text{grad}~{\mathbf{v}}+\left({\text{grad}~}{\mathbf{v}}\right)^{\top}\right],
\end{align}
where ${p}_\text{f}$ is the pressure and ${\eta}_\text{f}$ is the viscosity of the fluid. We assume the sphere is translating with velocity ${\mathbf{U}}$ thus the no slip boundary condition dictates ${\mathbf{v}}={\mathbf{U}}$ at the fluid-solid interface. In the solid domain ($\Omega_{\text{s}}$), the governing equations are described in terms of material coordinates. Thus, to avoid any confusion, we write the material gradient, divergence and Laplacian using ${\boldsymbol{\nabla}}$, $\boldsymbol{{\nabla}}\cdot$ and ${{\nabla}}^2$, respectively. The equilibrium momentum balance in $\Omega_{\text{s}}$ then yields
\begin{align}
\label{neo1}
&{\boldsymbol{\nabla}}\cdot{\boldsymbol{\sigma}}_\text{s} + {\mathbf{f}}(t)=\mathbf{0},
\end{align}
where ${\boldsymbol{\sigma}}_\text{s}$ is the solid elastic stress and $\mathbf{f}$ is a body force density on the sphere. Since the motion is axisymmetric, we assume the elastic sphere reaches a stable equilibrium, wherein the velocity gradient field in the solid domain is zero and the sphere has a rigid motion thereafter \cite{huang2011,villone2016}. As we will show later, for a weakly-elastic sphere, the leading-order effect of elasticity does not lead to any change in shape. Thus, a higher-order analysis is necessary in order to understand the change in shape of a translating elastic sphere. Extending linear elasticity to higher orders introduces further complexity by involving more material properties \cite{ogden1984}, instead here we use a phenomenological neo-Hookean model to capture the higher-order effects. The constitutive relation for an isotropic incompressible neo-Hookean solid can be expressed in terms of the displacement vector ${\mathbf{u}}$ \cite{gurtin2010,ogden1984} as
\begin{align}
{\boldsymbol{\sigma}}_\text{s}=-{p}_\text{s}\mathbf{{I}}  + {\eta}_\text{s}\left(\boldsymbol{\mathsf{F}}\cdot\boldsymbol{\mathsf{F}}^{\top}-\mathbf{{I}}\right),
\end{align}
where $\boldsymbol{\mathsf{F}}=\mathbf{{I}}+{\boldsymbol{\nabla}}{\mathbf{u}}$ is the deformation gradient tensor and ${\eta}_{\text{s}}$ is the shear modulus. For any material point, the displacement vector is defined $\mathbf{u}=\boldsymbol{\chi}\left(\mathbf{X},t \right)-\mathbf{X}$, where $\mathbf{X}$ is the position vector in the reference configuration (in other words material point) and $\boldsymbol{\chi}\left(\mathbf{X},t \right)$ is the deformation vector mapping each material point to its new location \cite{gurtin2010}. Here, ${p}_\text{s}$ serves only as a Lagrange multiplier to impose the incompressibility of the solid through
\begin{align}
\label{neo2}
&\text{det}(\boldsymbol{\mathsf{F}})=1,
\end{align}
where $\text{det}(\boldsymbol{\mathsf{F}})$ is the determinant of tensor $\boldsymbol{\mathsf{F}}$. The solid and fluid momentum balances are coupled through the continuity of normal traction at the interface ($\partial\Omega$), which dictates
\begin{align}
\label{continuity}
{\boldsymbol{\sigma}}_\text{s}\cdot\mathbf{n}={\boldsymbol{\sigma}}_\text{f}\cdot\mathbf{n},
\end{align}
where $\mathbf{n}$ is the normal vector to the surface of the deformed sphere.

Without any loss of generality, we will assume that the translational velocity, $\mathbf{U}=U\mathbf{e}_z$, and the body force density, $\mathbf{f}=bf(t)\mathbf{e}_z$, are oriented along $\mathbf{e}_z$. For simplicity we assume a spatially uniform body force where $b$ is a positive constant denoting the magnitude of the forcing while $f$ is a dimensionless $O(1)$ function such that the elastic deformation may be considered quasistatic.

Before going further, we non-dimensionalize all the equations defining dimensionless quantities $\hat{\boldsymbol{\nabla}}=R_0{\boldsymbol{\nabla}}$, $\hat{\mathbf{u}}={\mathbf{u}}/{R}_0$, $\hat{\mathbf{v}}={\mathbf{v}}/{U}_{\text{ch}}$, $\hat{\mathbf{U}}={\mathbf{U}}/{U}_{\text{ch}}$, $\hat{t}=t/(R_0/U_\text{ch})$, $\hat{p}_{\text{f}}={p}_{\text{f}}/({\eta}_{\text{f}}{U}_{\text{ch}}/{R}_0)$, $\hat{\boldsymbol{\sigma}}_{\text{f}}={\boldsymbol{\sigma}}_{\text{f}}/({\eta}_{\text{f}}{U}_{\text{ch}}/{R}_0)$, $\hat{\boldsymbol{\sigma}}_{\text{s}}={\boldsymbol{\sigma}}_{\text{s}}/{\eta}_{\text{s}}$, $\hat{p}_{\text{s}}={p}_{\text{s}}/{\eta}_{\text{s}}$ and $\hat{\mathbf{f}}(t)={\mathbf{f}}(t)/({\eta}_{\text{s}}/{R}_0)$, where ${U}_{\text{ch}}={2b {R}_0^2 }/{9{\eta}_{\text{f}}}$. Here ${U}_{\text{ch}}$ simply denotes the translational speed of a rigid sphere under a constant body force of magnitude $b$. Furthermore, for a forcing profile with frequency $\omega$, we define $\nu=\omega R_0/U_\text{ch}$ as a ratio of time scales. Now for convenience, we drop the ($~\hat{}~$) notation and henceforth refer to dimensionless variables. The dimensionless form of the boundary condition at the fluid-solid interface is then derived 
\begin{align}
\label{interface}
\boldsymbol{\sigma}_\text{s}\cdot\mathbf{n}=\epsilon\boldsymbol{\sigma}_\text{f}\cdot\mathbf{n},
\end{align}
where $\boldsymbol{\sigma}_\text{f} = -p_\text{f}{\mathbf{I}} + \text{grad}~\mathbf{v} + \left(\text{grad}~\mathbf{v}\right)^{\top}$ is the dimensionless stress in the fluid, $\boldsymbol{\sigma}_\text{s}= -p_\text{s}{\mathbf{I}} + \boldsymbol{\nabla}\mathbf{u} + \boldsymbol{\nabla}\mathbf{u}^{\top} + \boldsymbol{\nabla}\mathbf{u}\cdot\boldsymbol{\nabla}\mathbf{u}^{\top}$ is the dimensionless stress in the solid phase and $\epsilon={{\eta}_{\text{f}} {U}_\text{ch}}/{{\eta}_{\text{s}} {R}_0}$ represents the ratio of the viscous forces to the elastic forces. The relaxation time scale of the solid $\tau_\text{relax}\sim \eta_\text{f}/\eta_\text{s}$ which when non-dimensionlized scales as $O(\epsilon)$. Thus, for $\epsilon\ll 1$, the time required for relaxation is asymptotically shorter than the imposed time scale of motion, which justifies the quasistatic assumption.

In order to develop a geometric relation between the displacement vector and the surface deformation, we consider spherical coordinate systems $(r,\theta,\phi)$ in the spatial configuration. Since the motion is axisymmetric, we can define the surface as $r_\text{s}(\theta)$ where $\theta$ is the polar angle. Thus, at the interface, this definition yields
\begin{align}
\label{geo}
||\mathbf{X}+\mathbf{u}||=r_\text{s},
\end{align}
providing a geometric relation between surface equation and the displacement vector. We should emphasize that the governing equations in $\Omega_\text{s}$ are expressed in a material description. Thus, to obtain the deformation in the spatial variables, we transform the results of equation \eqref{geo}, using the mapping $\boldsymbol{\chi}$.

\subsection{Asymptotic analysis}
Here we focus on the case wherein the elastic forces are much larger than the viscous forces, i.e., $\epsilon\ll 1$. We expand all the parameters in terms of $\epsilon$ and refer to the $i^{\text{th}}$ order of any parameter using superscript $(i)$ (e.g., $p_\text{f}= p_\text{f}^{(0)}+\epsilon p_\text{f}^{(1)}+\epsilon^2 p_\text{f}^{(3)}+\cdots$). Due to the linearity of the Stokes equations, at any order the flow field around the sphere is governed by
\begin{align}
-&\text{grad}~p_\text{f}^{(i)}+ \text{div}\left(\text{grad}~\mathbf{v}^{(i)}\right)=\mathbf{0},\\
&\text{div}~\mathbf{v}^{(i)}=0,
\end{align}
where $\boldsymbol{\sigma}_\text{f} ^{(i)}= -p_\text{f}^{(i)}\mathbf{I} + \text{grad}~{\mathbf{v}}+\left({\text{grad}~}{\mathbf{v}}\right)^{\top}$ and $i\in\{0,1,2, \cdots\}$. We use the general solution given by Sampson for axisymmetric Stokes flow in the spherical coordinate system \cite{sampson1891,happel1981}. The boundary conditions in the fluid domain thereby are $\mathbf{v}^{(i)} =0$ at $r\rightarrow\infty$ and $\mathbf{v} = U\mathbf{e}_z$ at $r=r_\text{s}$. In the solid domain, the nonlinear governing equations are linearized perturbatively, thus we treat the problem at each order separately. As one can notice from equation \eqref{interface}, there is no deformation at the zeroth order thus the leading-order elastic effects are of $O(\epsilon)$. Throughout the following analysis we first solve the solid domain equations using a material description, and then map to the spatial configuration to enforce the interface boundary conditions. All formula given below for $\mathbf{u}$ and $p_\text{s}$ are reported in terms of spatial variables.
\subsubsection{Zeroth order flow field (first order solid deformations)}
At zeroth order in the fluid domain, the motion is simply the translation of a rigid sphere in Stokes flow. Satisfying $\mathbf{v}^{(0)}=f \mathbf{e}_z$ at $r=1$, we find 
\begin{align}
v_r^{(0)}&=\frac{{f}}{2}\left(\frac{3}{r}-\frac{1}{r^3}\right)\cos\theta,\\
v_\theta^{(0)}&=-\frac{{f}}{4}\left(\frac{3}{r}+\frac{1}{r^3}\right)\sin\theta,\\
p_\text{f}^{(0)}&=\frac{3{f}}{2r^2}\cos\theta.
\end{align}
The leading-order deformation equations in the solid domain are in the form of the Stokes equations as
\begin{align}
-&\boldsymbol{\nabla} p_\text{s}^{(1)}+ \nabla^2\mathbf{u}^{(1)}+ \mathbf{f}(t)=\mathbf{0},\\
&\boldsymbol{\nabla}\cdot\mathbf{u}^{(1)} =0,\\
&\boldsymbol{\sigma}_\text{s}^{(1)}= -p_\text{s}^{(1)}\mathbf{I} + \boldsymbol{\nabla}\mathbf{u}^{(1)}+ \boldsymbol{\nabla}\mathbf{u}^{{(1)}{\top}}.
\end{align}
Thus, here as well, we can employ Sampson's general solution for an axisymmetric Stokes flow. At this order, the interface boundary condition is $\sigma_{\text{s},rr}^{(1)}=\sigma_{\text{f},rr}^{(0)}$ and $\sigma_{\text{s},r\theta}^{(1)}=\sigma_{\text{f},r\theta}^{(0)}$. Noting that at this order the reference and spatial configurations coincide, we obtain
\begin{align}
u_r^{(1)}&=\frac{{f}}{2}(1-r^2)\cos\theta,\\
u_\theta^{(1)}&=\frac{{f}}{2}(-1+2r^2)\sin\theta,\\
p_\text{s}^{(1)}&=-\frac{{f}}{2}r\cos\theta.
\end{align}
To find the surface deformation, we define surface equation $r_\text{s}=1+s(\theta)$ and use the geometric relation in \eqref{geo}, which at this order leads to  $s^{(1)}=u_{r}^{(1)}$ at $r=1$. Therefore, we find $s^{(1)}=0$, indicating that the elastic sphere remains spherical with no surface deformation. We note that this result is similar to the sedimentation of a falling drop in a viscous fluid. \citet{Taylor1964} showed that  when inertia is neglected and the flow fields both inside and outside the drop are similarly governed by the Stokes equations, the shape has to remain spherical to satisfy the continuity of the normal tractions at the interface. 
\subsubsection{First-order flow field (second order solid deformations)}
At this order, the flow field at surface of the sphere satisfies $\mathbf{v}^{(1)}=U^{(1)} \mathbf{e}_z$. Recalling that $s^{(1)}=0$, we find
\begin{align}
\label{U1}
v_r^{(1)}&=\frac{U_1 {f}^2}{2}\left(\frac{3}{r}-\frac{1}{r^3}\right)\cos\theta,\\
v_\theta^{(1)}&=-\frac{U_1 {f}^2}{4}\left(\frac{3}{r}+\frac{1}{r^3}\right)\sin\theta,\\
p_\text{f}^{(1)}&=\frac{3U_1 {f}^2\cos\theta}{2r^2},
\end{align}
where the first correction for translational velocity $U_1$ shall be determined by satisfying the interface boundary condition. In the solid domain, the governing equations are
\begin{align}
-&\boldsymbol{\nabla} p_\text{s}^{(2)}+ \nabla^2\mathbf{u}^{(2)}+\boldsymbol{\nabla}\left(\boldsymbol{\nabla}\cdot\mathbf{u}^{(2)}\right)+ \boldsymbol{\nabla}\cdot\left( \boldsymbol{\nabla}\mathbf{u}^{(1)}\cdot\boldsymbol{\nabla}\mathbf{u}^{{(1)}{\top}}\right)=\mathbf{0},\\
&\boldsymbol{\nabla}\cdot\mathbf{u}^{(2)}+ \text{tr}(\boldsymbol{\nabla}\mathbf{u}^{(1),\text{c}}) =0,
\end{align}
where $\text{tr}(~)$ and $(~)^{\text{c}}$ indicate trace and cofactor of the tensor, respectively. Here the stress in the solid phase is defined
\begin{align}
&\boldsymbol{\sigma}_\text{s}^{(2)}= -p_\text{s}^{(2)}\mathbf{I} + \boldsymbol{\nabla}\mathbf{u}^{(2)}+ \boldsymbol{\nabla}\mathbf{u}^{{(2)}{\top}} +\boldsymbol{\nabla}\mathbf{u}^{(1)}\cdot\boldsymbol{\nabla}\mathbf{u}^{{(1)}{\top}}.
\end{align}
Now, by enforcing the interface boundary conditions $\sigma_{\text{s},rr}^{(2)}=\sigma_{\text{f},rr}^{(1)}$ and $\sigma_{\text{s},r\theta}^{(2)}=\sigma_{\text{f},r\theta}^{(1)}$ at $r=1$, we find $U_1=0$ and
\begin{align}
u_r^{(2)}&=- \frac{{f}^2 r}{304}\left(23+27r^2+(69+5 r^2)\cos2\theta\right),\\
u_\theta^{(2)}&=\frac{{f}^2 r}{304}\left(69+97r^2\right)\sin2\theta,\\
p_\text{s}^{(2)}&=\frac{{f}^2}{152}\left(190-199r^2-27r^2\cos2\theta\right),
\end{align}
leading to $p_\text{f}^{(1)}=0$, $v_r^{(1)}=0$ and $v_\theta^{(1)}=0$. It is worthwhile to emphasize that the leading order corrections for the flow field (i.e., $v_r^{(1)}$ and $v_\theta^{(1)}$) are imposed by the leading order deformation in the solid domain. Thus, $s^{(1)}=0$ indeed causes no disturbance in the flow field at this order. Finally, to find  $s^{(2)}$, we use the geometric relation
\begin{align}
s^{(2)}=u_r^{(2)} + \frac{\left(u_\theta^{(1)}\right)^2}{2},\quad\text{at}~r=1,
\end{align}
leading to $s^{(2)}=-\frac{31{f}^2}{304}(1+3\cos2\theta)$, which indicates a shape deviation from a sphere to an oblate spheroid of aspect ratio $1-\frac{93}{152}{f}^2\epsilon^2$.
\subsubsection{Second-order flow field (third-order solid deformations)}
\begin{figure}
\centering
\includegraphics[scale=0.35]{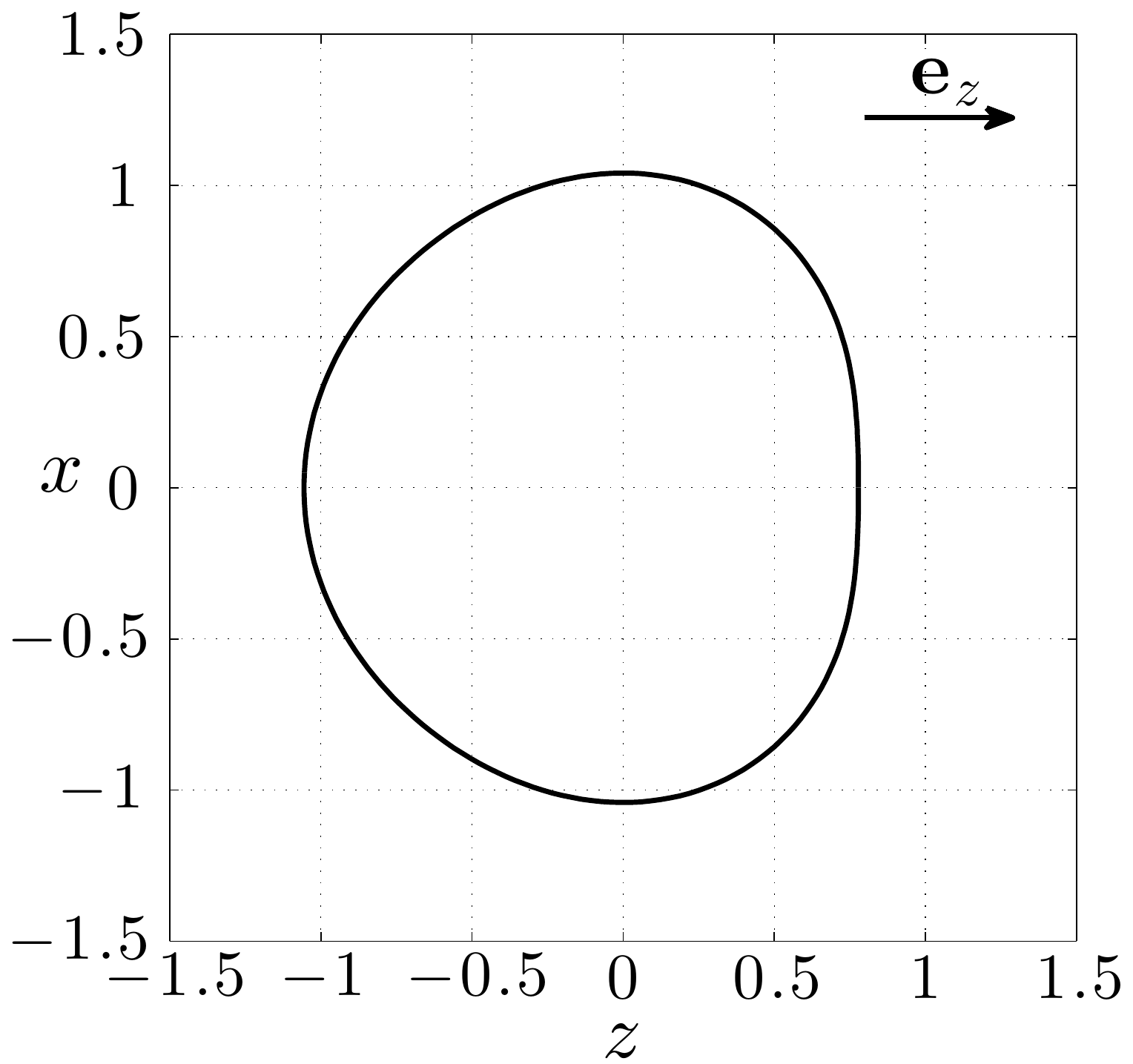} 
\caption{Deformed shape of the translating elastic sphere when $f=1$ and $\epsilon=0.45$.}
  \label{shape} 
\end{figure}

The no slip boundary condition for the Stokes equations at this order is $\mathbf{v}^{(2)}+s^{(2)}\frac{\partial \mathbf{v}^{(0)}}{\partial r}=U^{(2)} \mathbf{e}_z$. The second order flow field around the sphere is
\begin{align}
\label{U2}
v_r^{(2)}&= \frac{U_2 {f}^3 \cos\theta}{2}\left(\frac{3}{r}-\frac{1}{r^3} \right) +\frac{93{{f}^3}\cos\theta}{1520}\left( \frac{2}{r} +\frac{1-15\cos2\theta}{r^3} -\frac{3-15\cos2\theta}{r^5} \right),\\
v_\theta^{(2)}&= -\frac{U_2{{f}^3}\sin\theta}{4}\left(\frac{3}{r}+\frac{1}{r^3} \right)-\frac{93{{f}^3}\sin\theta}{6080}\left( \frac{4}{r} +\frac{13+15\cos2\theta}{r^3} -\frac{27+45\cos2\theta}{r^5} \right),\\
p_\text{f}^{(2)}&=\frac{3 U_2{{f}^3} \cos\theta}{2 r^2}-\frac{93{{f}^3}\cos\theta}{3040}\left( \frac{4}{r^2} +\frac{15-75\cos2\theta}{r^4}  \right).
\end{align}
Similar to the previous order, to determine the correction for the translational velocity (i.e., $U_2$), we need to solve the solid deformation equations at the third order given by
\begin{align}
-&\boldsymbol{\nabla} p_\text{s}^{(3)}+ \nabla^2\mathbf{u}^{(3)}+\boldsymbol{\nabla}\left(\boldsymbol{\nabla}\cdot\mathbf{u}^{(3)}\right)+ \boldsymbol{\nabla}\cdot\left( \boldsymbol{\nabla}\mathbf{u}^{(1)}\cdot\boldsymbol{\nabla}\mathbf{u}^{{(2)}{\top}}+\boldsymbol{\nabla}\mathbf{u}^{(2)}\cdot\boldsymbol{\nabla}\mathbf{u}^{{(1)}{\top}}\right)=\mathbf{0},\\
&\boldsymbol{\nabla}\cdot\mathbf{u}^{(3)}+\text{det}\left(  \boldsymbol{\nabla}\mathbf{u}^{(1)}\right)+{T}=0,
\end{align}
where ${T}$ is the $O(\epsilon^3)$ contribution of $\text{tr}(\boldsymbol{\nabla}\mathbf{u}^{\text{c}})$. Here, the stress inside the solid is defined
\begin{align}
&\boldsymbol{\sigma}_\text{s}^{(3)}= -p_\text{s}^{(3)}\mathbf{I} + \boldsymbol{\nabla}\mathbf{u}^{(3)}+ \boldsymbol{\nabla}\mathbf{u}^{{(3)}{\top}} + \boldsymbol{\nabla}\mathbf{u}^{(1)}\cdot\boldsymbol{\nabla}\mathbf{u}^{{(2)}{\top}}+\boldsymbol{\nabla}\mathbf{u}^{(2)}\cdot\boldsymbol{\nabla}\mathbf{u}^{{(1)}{\top}}.
\end{align}
Enforcing the third order interface boundary conditions at $r=1$ as
\begin{align}
\sigma_{\text{s},rr}^{(3)}+s^{(2)}\frac{\partial \sigma_{\text{s},rr}^{(1)}}{\partial r}-\frac{\text{d}s^{(2)}}{\text{d}\theta}\sigma_{\text{s},r\theta}^{(1)}&=\sigma_{\text{f},rr}^{(2)}+s^{(2)}\frac{\partial \sigma_{\text{f},rr}^{(0)}}{\partial r}-\frac{\text{d}s^{(2)}}{\text{d}\theta}\sigma_{\text{f},r\theta}^{(0)},\\
\sigma_{\text{s},r\theta}^{(3)}+s^{(2)}\frac{\partial \sigma_{\text{s},r\theta}^{(1)}}{\partial r}-\frac{\text{d}s^{(2)}}{\text{d}\theta}\sigma_{\text{s},\theta\theta}^{(1)}&=\sigma_{\text{f},r\theta}^{(2)}+s^{(2)}\frac{\partial \sigma_{\text{f},r\theta}^{(0)}}{\partial r}-\frac{\text{d}s^{(2)}}{\text{d}\theta}\sigma_{\text{f},\theta\theta}^{(0)},
\end{align}
we finally find $U_2=-\frac{31{{f}^3}}{380}$, and
\begin{align}
\label{ur3}
u_r^{(3)}&= \left(7r^2 \left(310 r^2-2091\right) \cos 2 \theta-5816 r^4+9645r^2-30240\right)\frac{{{f}^3}\cos\theta}{25536} ,\\
\label{ut3}
u_\theta^{(3)}&=\left(7r^2 \left(434 r^2+697\right) \cos 2 \theta+5284 r^4+381r^2+10080\right)\frac{{{f}^3}\sin\theta}{8512},\\
\label{p3}
p_\text{s}^{(3)}&=-{r}\left(322r^2\cos2\theta+3766r^2-2991 \right)\frac{{{f}^3}\cos\theta}{1596}.
\end{align}
Thence, we can determine the third order shape deformation using the geometric relation \eqref{geo}, which reads $s^{(3)}=u_r^{(3)}+u_\theta^{(1)}u_\theta^{(2)}$ at $r=1$. Notably, we find $s^{(3)}=-\frac{2777 {{f}^3}}{1824}\cos^3\theta$ indicating a `egg-like' deformation which exhibits a front-back asymmetry in the surface of the elastic sphere as shown in figure~\ref{shape}. To recover the solution for the case of translation under a constant body force (i.e. sedimentation), one can set $f=1$. Then  $U={1}-\frac{31}{380}\epsilon^2+O(\epsilon^3)$ suggesting a slower translational velocity, which is notably unlike the settling speed of a compressible Hookean sphere \cite{murata1980}. It is also worthwhile to note that elastic capsules containing viscous fluids exhibit a similar asymmetry in their deformation under pure translation. In a numerical study, \citet{ishikawa2016} showed that at steady state, a weakly-elastic spherical micro-torque swimmer deforms to an egg-like shape. A similar deformation was observed experimentally for sedimenting vesicles as well \cite{huang2011}.
\subsubsection{Third-order flow field}
To quantify the effect of the shape asymmetry on the motion of the particle, we shall determine the third order correction for the flow field. Once again, we solve the Stokes equations, but this time with $\mathbf{v}^{(3)}+s^{(3)}\frac{\partial \mathbf{v}^{(0)}}{\partial r}=U^{(3)} \mathbf{e}_z$ at $r=1$. Thus, we can find the third order correction for the fluid field and stress field in terms of the translational velocity $U^{(3)}$. Now to find $U^{(3)}$, instead of solving for the next order solid deformation (as we did in the previous orders), we employ an auxiliary case wherein a rigid sphere of the same radius is translating with the same driving force \cite{payne1960}. Since the motion is over-damped, regardless of the shape, the driving force is always entirely balanced by the viscous surface forces. Thus, for a given driving force, the net drag force on both elastic and rigid spheres are the same. For the elastic sphere we have
\begin{align}
\mathbf{F}^{\text{dr}}+3\epsilon\int_{\partial\Omega}\boldsymbol{\sigma}_{\text{f}}\cdot\mathbf{n}\text{d}S=\mathbf{0},
\end{align}
where $\mathbf{F}^{\text{dr}}=\mathbf{f} V$ is the total driving force on the sphere, $V$ is the volume and $S$ represent dimensionless area element of the sphere. On the other hand, for the rigid sphere case, the drag law dictates $\mathbf{F}^{\text{dr}}/V=\frac{9{{f}}}{2}\epsilon~\mathbf{e}_z$, thus
\begin{align}
\label{recip}
\frac{1}{V}\int_{\partial\Omega}\boldsymbol{\sigma}_{\text{f}}\cdot\mathbf{n}\text{d}S=-\frac{3{{f}}}{2}\mathbf{e}_z.
\end{align}
Now by substituting $\boldsymbol{\sigma}_{\text{f}}=\boldsymbol{\sigma}^{(0)}_{\text{f}}+\epsilon\boldsymbol{\sigma}^{(1)}_{\text{f}}+\epsilon^2\boldsymbol{\sigma}^{(2)}_{\text{f}}+\epsilon^3\boldsymbol{\sigma}^{(3)}_{\text{f}}$, we determine the left-side of equation \eqref{recip} as $\frac{1}{V}\int_{\partial\Omega}\boldsymbol{\sigma}_{\text{f}}\cdot\mathbf{n}\text{d}S=-\frac{3}{2}\left({{f}}+U^{(3)}\epsilon^3\right)\mathbf{e}_z$, indicating that $U^{(3)}=0$. Thus, the final expression for the translational velocity 
\begin{align}
U=\left[1-\frac{31}{380}f^2\epsilon^2 +{O}(\epsilon^4)\right]f,
\end{align}
and the third order corrections in the flow field are
\begin{align}
v_r^{(3)}&=\frac{2777{{f}^4}}{17024}\left(\frac{1}{r^2}-\frac{1}{r^4}\right)\left( 3-9\cos^2\theta -\frac{3-30\cos^2\theta+35\cos^4\theta}{r^2} \right),\\
v_\theta^{(3)}&=\frac{2777{{f}^4}}{34048}\left(\frac{\sin2\theta}{r^4}\right)\left( {12}-14\cos^2\theta-\frac{12-28\cos^2\theta}{r^2}    \right),\\
p_\text{f}^{(3)}&=\frac{2777{{f}^4}}{42560}\left(\frac{1}{r^4}\right)\left( 15-45\cos^2\theta -\frac{21-210\cos^2\theta+245\cos^4\theta}{r^2}  \right).
\end{align}
\section{Two-sphere swimmer}
\label{two-swimmer}
\begin{figure}
\centering
\includegraphics[scale=0.45]{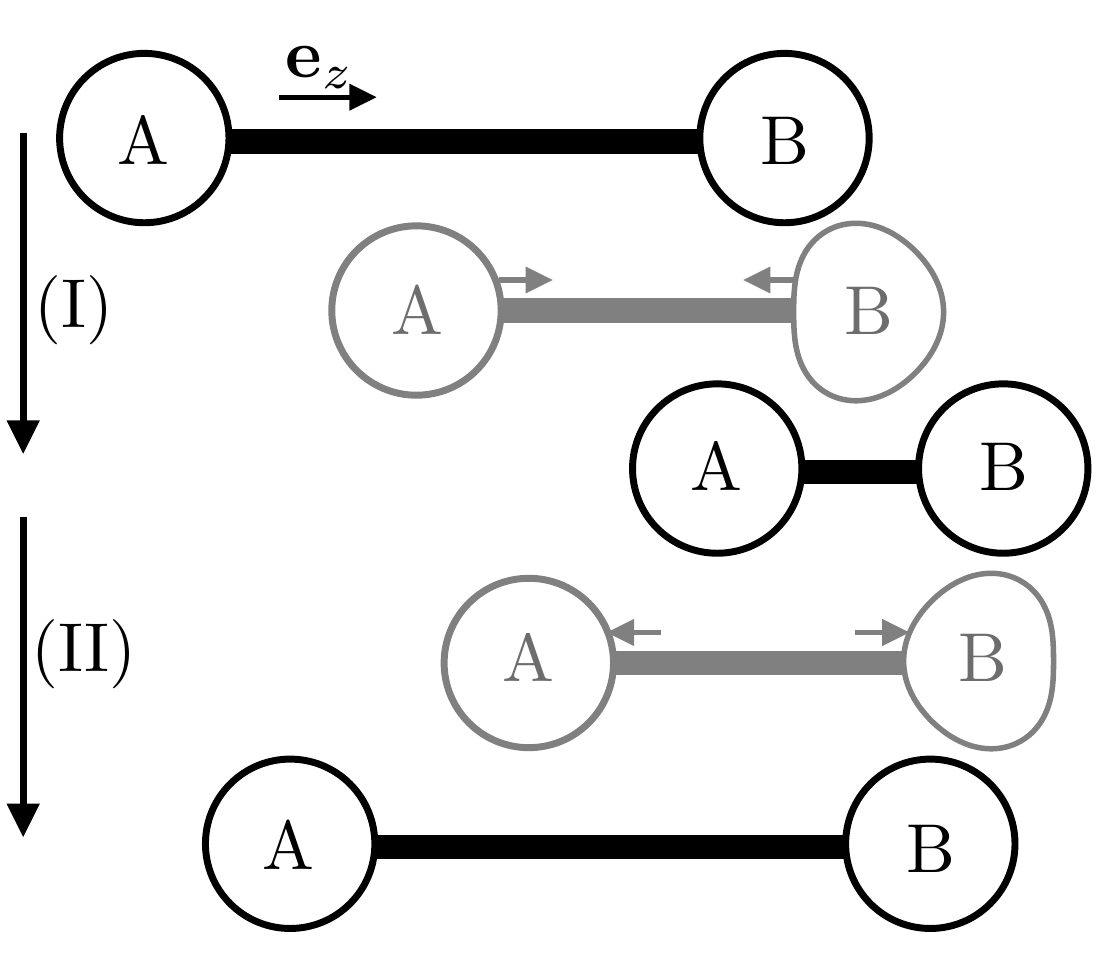}
\caption{One cycle of the two-step motion of the swimmer. Step (I): The rod shortens its length. Step (II): Spheres move away from one another until they reach the initial distance. The steps in grey colour demonstrate the swimmer while it proceeds to the next step and sphere B is deformed.}
\label{swimmer}
\end{figure}

We consider a model swimmer which consists of two spheres: a rigid sphere A and a neo-Hookean isotropic incompressible elastic sphere B (identical to the elastic sphere defined in section~\ref{single-sphere}). The spheres are of equal radii and linked by a rod of length $L$. To propel itself forward, the swimmer repeats a two-step, one-dimensional motion in which the connecting rod shortens its length in step (I), and then returns back to its original length in step (II) in a harmonic fashion (see figure~\ref{swimmer}). While advancing from one step to another, sphere B changes its shape continuously and instantaneously, until it reaches its spherical shape again at the end point of each step. We note that despite the reversible actuation, the flow field induced by sphere B is not front-back symmetric. Thus, for sphere A, the contribution of the background flow (induced by sphere B) is different between step (I) and (II). The net motion in each cycle thereby is not kinematically reversible and indeed the swimmer can propel with a velocity that we determine below.

The connecting rod exerts driving forces $\mathbf{F}_\text{A}$ and $\mathbf{F}_\text{B}$ on spheres A and B, respectively. The force-free motion of the swimmer necessitates $\mathbf{F}_\text{A}+\mathbf{F}_\text{B}=\mathbf{0}$. Although, in practice, the driving forces are applied locally at the sphere-rod junctions, here we neglect the effect of rod and assume a spatially uniform force density for both spheres, noting that such actuation forces can be imposed by magnetic fields or optical tweezerss. Thus, we prescribe the periodic motions by defining  $\mathbf{F}_\text{A}/V=-\mathbf{F}_\text{B}/V=\frac{9}{2}\epsilon  \sin(\nu t)\mathbf{e}_z$. Assuming that spheres are well separated at all times, we employ a far-field approximation to determine the flow field around the swimmer. The velocity of each sphere, i.e. $\mathbf{U}_\text{A}$ and $\mathbf{U}_\text{B}$, then follows the drag law
\begin{align}
\label{RTD1}
\mathbf{U}_\text{A}=\mathbf{R}_\text{A}^{-1}\cdot\mathbf{F}_\text{A}+\mathcal{F}_\text{A}\left[\mathbf{{v}}^{}_{\text{B}\rightarrow\text{A}}\right],\\
\label{RTD2}
\mathbf{U}_\text{B}=\mathbf{R}_\text{B}^{-1}\cdot\mathbf{F}_\text{B}+\mathcal{F}_\text{B}\left[\mathbf{v}^{}_{\text{A}\rightarrow\text{B}}\right],
\end{align}
where $\mathbf{R}_\text{A}$ and $\mathbf{R}_\text{B}$ are hydrodynamic resistance tensors for spheres A and B, $\mathcal{F}_\text{A}$ and $\mathcal{F}_\text{B}$ are the Fax\`en operators, and $\mathbf{{v}}^{}_{\text{B}\rightarrow\text{A}}$ ($\mathbf{{v}}^{}_{\text{A}\rightarrow\text{B}}$) is the background flow field on sphere A (B), induced by sphere B (A). Here, to focus only on the leading order propulsion velocity, we limit our calculations to the first reflection of the flow fields. Therefore, we neglect the contribution of the background flow on the deformation of sphere B. At each step, we take the average velocity of the two spheres as the instantaneous velocity of the swimmer, defining $\mathbf{U}^{\text{(I)}}=\frac{\mathbf{U}_\text{A}^{\text{(I)}}+\mathbf{U}_\text{B}^{\text{(I)}}}{2}$  and $\mathbf{U}^{\text{(II)}}=\frac{\mathbf{U}_\text{A}^{\text{(II)}}+\mathbf{U}_\text{B}^{\text{(II)}}}{2}$, where superscripts (I) and (II) refer to the quantities at the corresponding steps. Thence, to find the net propulsion velocity we average the swimming velocities over one complete cycle
\begin{align}
\label{speed}
\bar{\mathbf{U}}=\frac{1}{\tau}\left(\int_{0}^{\tau/2}\mathbf{U}^{\text{(I)}}\text{d}t + \int_{\tau/2}^{\tau}\mathbf{U}^{\text{(II)}}\text{d}t\right),
\end{align}
where $\tau=2\pi/\nu$ is the period of the cycle. By making use of equations \eqref{RTD1} and \eqref{RTD2} and noting that $\mathbf{F}_\text{A} (t+\frac{\tau}{2})=-\mathbf{F}_\text{A} (t) $ and $\mathbf{F}_\text{B} (t+\frac{\tau}{2})=-\mathbf{F}_\text{B} (t) $, equation \eqref{speed} can be reduced to
\begin{align}
\label{integration}
\bar{\mathbf{U}}=\frac{1}{2\tau}\left\{\int_{0}^{\tau/2}\left(  \mathcal{F}_\text{A}\left[\mathbf{{v}}^{\text{(I)}}_{\text{B}\rightarrow\text{A}}\right] + \mathcal{F}_\text{B}\left[ \mathbf{{v}}^{\text{(I)}}_{\text{A}\rightarrow\text{B}} \right]\right)\text{d}t+ \int_{\tau/2}^{\tau}\left( \mathcal{F}_\text{A}\left[\mathbf{{v}}^{\text{(II)}}_{\text{B}\rightarrow\text{A}} \right]+ \mathcal{F}_\text{B}\left[ \mathbf{{v}}^{\text{(II)}}_{\text{A}\rightarrow\text{B}}  \right]\right)  \text{d}t\right\}.
\end{align}
From the general description of Fax\`en operator \cite{brenner1964,kim1985}, we find $\mathcal{F}_\text{A}=1+\frac{1}{6}\nabla^2$ and $\mathcal{F}_\text{B}=1+\left(\frac{1}{6}+\frac{31}{228}\epsilon^2\right)\nabla^2+O(\epsilon^3/l^2)$. Now using the asymptotic descriptions of the flow fields reported in section~\ref{single-sphere}, we arrive at the leading order propulsion velocity
\begin{align}
\bar{\mathbf{U}}=\frac{24993}{136192}\frac{\epsilon^3 }{L^2}\mathbf{e}_z.
\end{align}
We note that the deformation of the sphere governs the propulsive thrust and that the magnitude of the change in distance between the spheres does not contribute to the leading order motion, unlike the three sphere swimmer where the difference in arm lengths quantifies the asymmetry \cite{najafi2004}.

\section{Conclusion} 
In this paper, we inquired about the effects of elasticity on swimming in Stokes flow. We started by addressing the pure translation of an elastic particle in viscous fluid. We asymptotically showed that under a body force the translational velocity of an elastic sphere is slower, and also the shape deformation is not front-back symmetric. The latter indicates an asymmetry in the surrounding flow field which can be exploited to evade the scallop theorem. To highlight the effect of this deformation on swimming, we proposed a very simple swimmer of two spheres that can swim with a reversible actuation, solely due to elasticity of one of the spheres. Our results show that accounting for elasticity of bodies may be crucial to fully understand the dynamics of swimming cells and specifically can be useful in designing microswimmers. Finally we note that while conceptually simple, our elastic two-sphere swimmer is not very effective for small deformations, but in practice one might use an elastic body which is already asymmetric to exacerbate this effect.

\begin{acknowledgements}
The authors thank Professor G. M. Homsy for helpful discussions and support to B.N. through NSERC Grant No. RGPIN-386202-10. G.J.E. acknowledges funding from the NSERC Grant No. RGPIN-2014-06577.
\end{acknowledgements}

\bibliography{reference}

\end{document}